# Phonon-Polariton Propagation, Guidance, and Control in Bulk and Patterned Thin Film Ferroelectric Crystals

David W. Ward, Eric Statz, Jaime D. Beers, Nikolay Stoyanov, Thomas Feurer, Ryan M. Roth[1], Richard M. Osgood[1], and Keith A. Nelson
The Massachusetts Institute of Technology,
Cambridge, MA 02139, USA
[1]Microelectronics Sciences Laboratories, Columbia University,
New York, New York 10027, USA

**ABSTRACT**

Using time resolved ultrafast spectroscopy, we have demonstrated that the far infrared (FIR) excitations in ferroelectric crystals may be modified through an arsenal of control techniques from the fields of guided waves, geometrical and Fourier optics, and optical pulse shaping. We show that $LiNbO_3$ and $LiTaO_3$ crystals of 10-250 μm thickness behave as slab waveguides for phonon-polaritons, which are admixtures of electromagnetic waves and lattice vibrations, when the polariton wavelength is on the order of or greater than the crystal thickness. Furthermore, we show that ferroelectric crystals are amenable to processing by ultrafast laser ablation, allowing for milling of user-defined patterns designed for guidance and control of phonon-polariton propagation. We have fabricated several functional structures including THz rectangular waveguides, resonators, splitters/couplers, interferometers, focusing reflectors, and diffractive elements. Electric field enhancement has been obtained with the reflective structures, through spatial shaping, of the optical excitation beam used for phonon-polariton generation, and through temporal pulse shaping to permit repetitive excitation of a phonon-polariton resonant cavity.

**INTRODUCTION**

At long wavelengths, the coupling of electromagnetic radiation to an optic phonon mode gives rise to a propagating excitation known as a phonon-polariton, henceforth referred to as a polariton, which displays the properties of both its phonon and photon constituents [1]. The exploitation of this dual-nature excitation is called polaritonics, in analogy to photonics. The phonon-polariton dispersion relation splits into two branches. The lower branch displays light-like dispersion for small wavevectors and evolves into phonon-like dispersion at higher wavevectors. The converse is true for the upper branch; however, the limited time resolution of our probe pulse precludes the resolution of upper branch polaritons. Exploitation of polariton electromagnetic wave character provides a direct channel of control over propagation and dispersion of not only electromagnetic but also lattice vibrational coherence and energy.

Coherent polaritons are generated by ultrafast optical pulses through impulsive stimulated Raman scattering (ISRS) [2]. The ability to spatially and temporally shape the optical excitation beam provides a channel of coherent control through which polariton spatial and temporal profiles may be specified.

**EXPERIMENTAL DETAILS**

We utilize two lasers for experiments and fabrication: a home-built Ti:sapphire multi-pass amplifier (800 nm, 50 fs, 1 KHz rep. rate, 700 μJ/pulse) seeded by a KM Labs oscillator (790 nm, 15 fs, 88 MHz rep. rate, 3 nj/pulse), and a Coherent RegA Ti:sapphire regenerative amplifier (800 nm, 200 fs, 250 KHz rep. rate, 6 μJ/pulse) seeded by a Coherent Mira oscillator.

Stoichiometric, X- or Y-cut, poled LiNbO$_3$ and LiTaO$_3$ crystals provide excellent conversion of optical light to polaritons due to their high electro-optic coefficients. In the experiments reported here, several types of samples were used. The thinnest samples, ~10 μm thick LiNbO$_3$, were fabricated by crystal ion slicing, which uses high-energy ion implantation at 3.8 MeV, combined with chemical etching, to exfoliate single-crystal sheets of metal oxide crystals [3]. 250 and 500 μm thick LiNbO$_3$ crystals were used for the patterned materials experiments, a 2 mm thick LiTaO$_3$ crystal with elliptically polished ends (curvature perpendicular to the optic axis) was used for large aperture focusing, and a 5 mm thick LiTaO$_3$ crystal was used for axicon experiments. All crystals are commercially manufactured with the exception of the 10 μm films.

In order to generate narrowband polariton waveforms of wavelength $\lambda$, a spatially periodic optical intensity pattern is projected onto the sample. The excitation beam is passed through a binary phase mask, and the $\pm 1$ orders of diffraction are overlapped at the sample to form an interference pattern of specified period $\lambda$ [4]. Polariton frequencies from ~0.1-7.5 THz with bandwidths on the order of 10 GHz are accessible with this setup.

The ionic displacements concomitant with polariton propagation modulate the index of refraction of the crystal, creating a unique time-dependent phase pattern that corresponds to the polariton amplitude distribution over the spatial extent of the crystal. Polariton imaging uses well known techniques of phase to amplitude conversion to image this phase pattern onto a CCD [5]. Images are acquired at different delays of an ultrafast probe pulse with respect to the pump, and the complete set of sequential images over a large range of delays constitutes a 'movie' that captures the propagating THz wave as a function of time [6]. Other probe techniques, including interferometric measurement, are used to monitor polariton propagation through a single point.

Patterning of a material with 10-20 μm lateral resolution is achieved by focusing a series of ultrafast laser pulses (100-200 μJ/pulse) onto a crystal using a microscope objective (NA~1.4), leaving a material void in the irradiated region. A computer actuated Burleigh 3-axis translation stage changes the position of the sample in the beam path in steps of 10 μm, allowing a user specified pattern to be cut into the sample with little to no user intervention [7].

Temporal shaping of the pump beam is achieved using the Deathstar pulseshaper, which consists of a system of two retroreflectors that divide a single ultrafast laser pulse into seven pulses that are evenly spaced temporally with repetition rate in the 5-1000 GHz range and whose intensities form a roughly Gaussian profile [8]. The resulting pulsetrain was focused to a 120 μm spot size at a polariton resonator structure. For the experiments reported here, pulsetrain repetition was tuned from 0.1 to 0.4 THz in 10 GHz steps to explore the frequency-dependent resonator response

## RESULTS AND DISCUSSION

### Polariton Slab Waveguide

Narrowband polariton generation and polariton imaging with a 400 nm wavelength probe were used to characterize the dispersive properties of a 10 μm thick slab of LiNbO$_3$ over a 6-100 μm range of polariton wavelengths. Figure 1a shows a typical CCD image of polaritons with 56 μm wavelength collected 5 ps after generation by crossed excitation pulses. The images at different probe delay times show the movement of the polariton peaks and nulls, and analysis of the set of results permits direct determination of the phase velocity. Since the images are redundant in the vertical dimension, signal-to-noise was enhanced by over an order of magnitude by integration and compression of each image along that dimension. The compressed images were rotated by 90 degrees and arranged in time order to generate a space vs. time graph as shown in figure 1b. Fourier filtering and examination of the spectral content for each polariton wavelength yielded the dispersion relation shown in figure 1c.

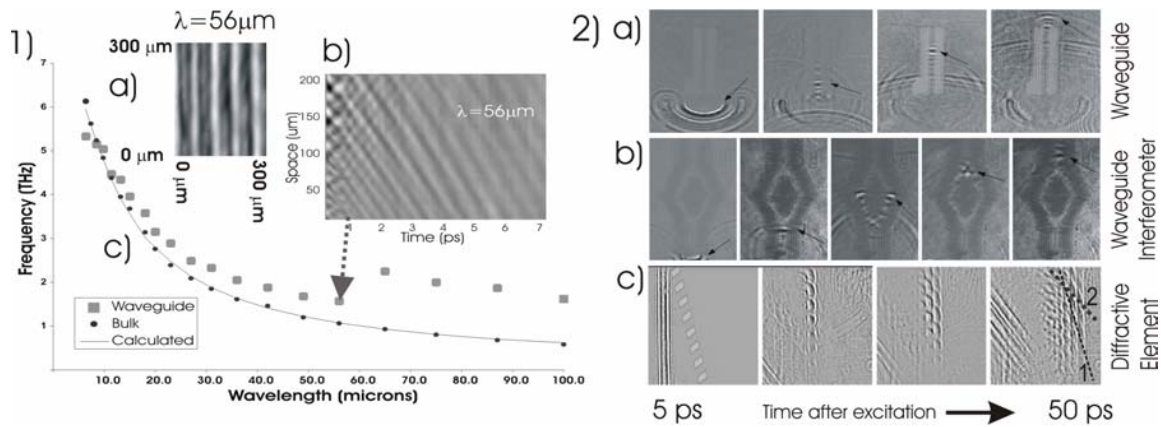

**Figure 1:** Crossed excitation beams generate narrowband polaritons in LiNbO$_3$, and their spatial and temporal evolution are monitored simultaneously by polariton imaging. a) CCD image of polaritons (56 μm wavelength) in a 10 μm thick LiNbO$_3$ film 5 ps after excitation. b) Space-time plot formed by compression and time ordering of images like that in (a), showing polariton propagation vs time. c) Polariton dispersion in a 10 μm thick film (squares) and bulk (circles and solid curve). **Figure 2:** Polariton propagation in several patterned materials demonstrating guidance, interference, and diffraction. a) 200 μm x 1.7 mm waveguide demonstrating polariton guidance. b) waveguide interferometer. c) Diffractive element consisting of ten 140 μm x 300 μm slits. The first two orders of diffraction are evident.

Comparison of the 10 μm slab waveguide to bulk LiNbO$_3$ in the limit of small polariton wavelength indicates similar dispersion properties. As the wavelength is increased, the dispersive properties deviate toward higher frequency than the corresponding bulk response, presumably under the influence of the lower-index exterior (air). At wavelengths much greater than the slab thickness, dispersion appears to be dominated by that of the exterior and has a group velocity very near that in air. Intermediate to these regimes, around 60 μm, the slab appears to undergo a rather sharp transition from predominantly bulk to predominantly exterior (i.e. "cladding" into which the polariton electromagnetic field extends) dispersion. We believe the abruptness of the transition may be due to the high dielectric contrast, 4.6 to 1.0. Assuming the transition is continuous, we expect to see a reversal of the group velocity over a short range of wavelengths in this region, but further investigation of this range is required.

## Patterned Samples

Through ultrafast laser machining, we have fabricated a variety of components for polariton guidance and passive signal processing. The images of polaritons propagating through a 1700x200 μm waveguide in figure 2a illustrates direct visualization of polariton confinement in a manner that is not generally possible for optical waveguides [7]. As the polaritons travel the extent of the waveguide, the wave front remains planar. Only upon exiting does it start to diverge and radiate as from a point source. Polariton propagation through a waveguide interferometer, as shown in figure 2b, demonstrates the potential for functional THz devices.

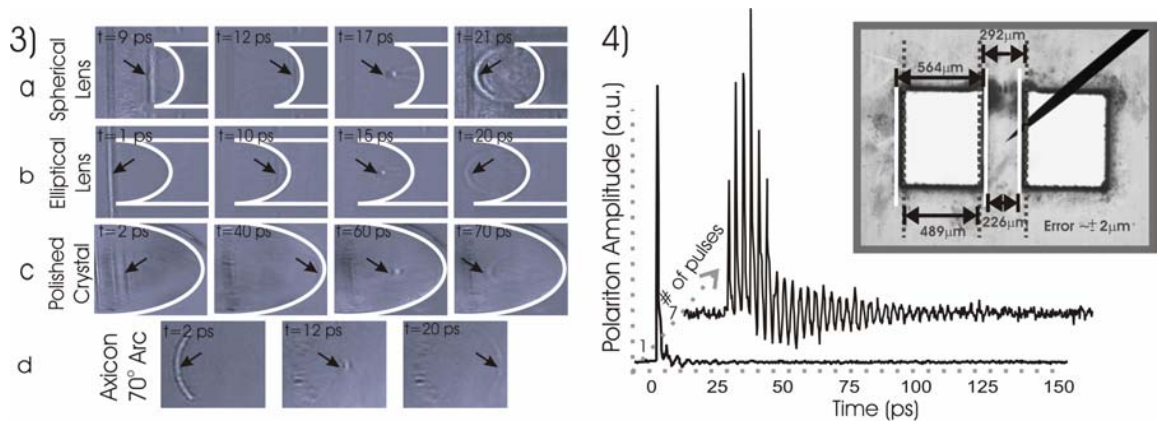

**Figure 3** Polariton electric field enhancement of ~2 via focusing with a) a laser machined LiNbO$_3$ spherical lens, b) a laser machined elliptical lens (major axis 300 µm, minor axis 294 µm), c) and a LiTaO$_3$ crystal with elliptically polished ends (major axis 2778 µm, minor axis 2500 µm), d) and enhancement of ~5 with a spatially shaped pump pulse (axicon focus) in LiTaO$_3$. **Figure 4** The LiNbO$_3$ region of the polariton resonator is indicated by the pointer and is flanked with hollow (air) cavities on either side. The combined effect of the two cavities is an enhanced resonance at 321 GHz and a slightly weaker resonance at 239 GHz. Two interferometric time scans of the polariton signal in the resonator are shown. The first shows the response generated by a single 1µj optical pulse within the LiNbO3 region and the second shows the enhanced response (~6x in intensity) from a series of seven optical pulses generated by the Deathstar pulseshaper tuned to the 321 GHz resonance.

The diffraction grating in figure 2c illustrates the potential for conducting polaritonics in the frequency domain [9]. The figure illustrates first and second order diffraction of a broadband THz pulse centered around 150 µm. The grating consists of ten 140x300 µm slits spaced 200 µm apart in a LiNbO$_3$ crystal. Such structures allow for spectral discrimination along a spatial dimension and have potential applications in THz spectroscopy.

Passive enhancement of polariton amplitudes has been realized through various methods of focusing. Through femtosecond laser machining, we have fabricated both spherical (figure 3a) and elliptical (figure 3b) focusing reflectors. We have also polished the end of a 3x5x2 mm LiTaO$_3$ crystal to an ellipse (figure 3c), forming a larger reflecting element. A similar focusing effect is achieved by spatially shaping the pump beam to a circular shape at the sample in order to generate a curved polariton wavefront which focuses as it propagates (figure 3d) [10,11]. We achieve this by imaging an arc from the output of an axicon (a conical biprism that produces a circular beam profile) onto the crystal. We achieved electric field enhancements of order 2-5 and intensity enhancements of 4-25 through these methods; however, theoretical estimates suggest that field enhancement of about 10 are achievable in both cases. Comparable additional enhancement could be achieved through focusing in the other transverse dimension.

## **Coherent Control in a Polariton Resonator**

We have used optical pulse shaping to repetitively drive polaritons trapped in a polariton resonator cavity (inset figure 4) with a primary resonance frequency of 321 GHz and an unloaded Q of approximately 10. There is also a weaker resonance at 239 GHz due to the coupling of hollow cavities to the central LiNbO$_3$ region as shown in the figure, but it is not explored in this report. Figure 4 presents interferometrically recorded data comparing the polariton response due to a single excitation pulse to that of seven pulses from the Deathstar pulseshaper. The strength and persistence of the polaritons generated by the pulsetrain are attributed to the hollow cavities flanking the resonator and will be discussed in detail in a subsequent publication.

We amplified polariton energy by a factor of 6.25 using the seven pulses from the Deathstar, which is 95% of the theoretical maximum obtainable with an infinite number of pulses. Propagation loss in the resonator is negligible, but transmission loss at the interface is high. The low Q of the cavity is the chief impediment to amplification. If the dielectric resonator's walls are coated with silver then the theoretical amplification maximum rises to near 1600, but the seven pulses from the Deathstar can only realize 15% of this. Thus, not only does the Q need to be increased to achieve near maximal amplification, but the number of pulses in the pulsetrain must also be increased.

**CONCLUSIONS**

Two channels of control over polariton propagation have been demonstrated. The first channel exerts control through manipulation of the electromagnetic wave character of the polaritons, which we have effected by fabricating polaritonic structures using ultrafast laser machining. The second acts through spatial or temporal shaping of the optical field which is passed on to the polariton response through the ISRS excitation mechanism. Control over polariton guidance, generation, propagation, dispersion, and frequency content has been shown. The THz radiation from polaritons may be propagated out of the crystal in which they are generated. Polariton control therefore enables a robust source of coherent THz radiation and a host of capabilities for THz signal processing and spectroscopy [12].

**ACKNOWLEDGEMENTS**

This work was supported in part by the National Science Foundation (CHE-0212375 and MRSEC Program, Grant No. DMR-0213282) and the U.S. Air Force Office of Scientific Research (contract # F49620-99-1-0038).